# Transparent and conductive graphene oxide-polyethylenglycol diacrylate coatings obtained by photopolymerization

M. Sangermano[1*], S. Marchi[1], L. Valentini[2**], S. Bittolo Bon[2], P. Fabbri[3]

[1] Politecnico di Torino, Dipartimento di Scienza dei Materiali e Ingegneria Chimica, Corso Duca degli Abruzzi 24, 10129 Torino, Italy.

[2] Università di Perugia, Dipartimento di Ingegneria Civile e Ambientale, Strada di Pentima 4, INSTM, UdR Perugia, 05100 Terni, Italy.

[3] Università di Modena e Reggio Emilia, Dipartimento di Ingegneria dei Materiali e dell'Ambiente, Via Vignolese 905/A, 41125, Modena, Italy.

**ABSTRACT**: Water dispersed graphene oxide sheets were used to prepare graphene-polyethylenglycol diacrylate resin composites by photopolymerization. It was found that graphene sheets undergo excellent morphological distribution within the resin system, giving rise to transparent composites with unaltered thermal properties with respect to the neat resin, that are electrically conductive at loading ratios as low as 0.02 %wt of graphene oxide . The proposed strategy based on photopolymerization provides an easy, energy-saving and environmental friendly technique that can find a wide application in coating technology, mainly for electromagnetic shielding and antistatic coatings.

Corresponding authors:
(*) marco.sangermano@polito.it
(**) mic@unipg.it

**Introduction**

Graphene, a new class of two-dimensional sp2 carbon nanostructure, holds great promise for potential applications in many technological fields such as optoelectronics. [1] One of the most common preparation method of graphene starts from the oxidation and exfoliation of graphite oxide to graphene oxide (GO). [2-4] Graphene oxide sheets consists of graphene sheet chemically functionalized with hydroxyl and epoxy groups, carbonyl groups are also present as carboxylic acids along the sheet edges. [5,6]

These functional groups should be purposely exploited to develop GO-based materials. For example, GO is intrinsically hydrophilic and its dispersion in water is easily achieved. [7-10] Graphene sheets were demonstrated to have extraordinary electronic transport properties, [11-15] combined with a wide set of other unusual properties: [16] their thermal conductivity and mechanical stiffness may rival the remarkable in-plane values for graphite (3,000$Wm^{-1}K^{-1}$ and 1,060 GPa, respectively); their fracture strength should be comparable to that of carbon nanotubes for similar types of defects. [17,18]

On the basis of these examplary physico-mechanical properties, it results clear that the transfer of such features to polymeric materials suitable for real applications, holds a deep scientific interest. However, for the time being, just a very limited number of studies have been done on the harnessing of these properties into functional graphene-based polymer composites: an important step was actually done in understanding that graphene-based polymer composites exhibit extraordinarily low electrical percolation threshold (0.1 vol %) due to the large conductivity and aspect ratio of graphene sheets,[19,20] thus suggesting a possible introduction of percolating graphene network within an insulating material rendering it a semiconductor. Although several graphene nanocomposites have been described so far, the achievement of an uniform dispersion of graphene nanosheets into a polymer matrix is still a challenge and deserves a deep and intense investigation. Because of their high specific surface area and surface chemical features, graphene sheets tend to be arranged in agglomerates or stacks due to Van der Waals interactions. The complete exploitation of the intrinsic potential of graphene nanosheets will inevitably depend on the possibility to achieve a good degree of exfoliation and dispersion in the polymer matrix. The vast majority of the strategies proposed to achieve highly dispersed graphene in polymer matrices often goes through complex surfacial chemistry routes, such as the covalent functionalization of graphene via ester linkages [21], amide linkages [22,23], and others. Besides the complex chemistry, covalent attachment to graphene may hold a significant drawback in some instances, as the covalent bonds have the potential to disrupt the conjugated structure, leading to compromised physical properties. Therefore, an alternative to covalent bonding could have significant potential advantages for real applications, and physical dispersion of graphene into polymer precursors still represents the easiest route.

In this view, the preparation of polymer composites through UV curing presents several advantages, mainly related to the easiness of dispersing GO nanosheets into the liquid, low-viscosity monomers before curing. Actually, the UV curing technique is getting an increasing importance in the field of coatings due to its peculiar characteristics:[24] it induces the polymer formation with a fast transformation of the liquid monomer into a solid film with tailored physical-chemical and mechanical properties; it can be considered an environmental friendly technique, being solvent free, and it is usually carried out at room temperature, with energy saving.

Despite of the promising application of GO in UV curable resins, to the best of our knowledge no reports are present in the literature so far. Therefore this paper intends to investigate the effect of GO on the UV-curing process of an acrylic resin of widely diffused application, namely polyethylenglycol diacrylate, and to study the electrical conductivity of the cured films. The proposed strategy based on photopolymerization provides an easy, energy-saving and environment friendly technique that can find a wide application in coatings technology, mainly for electromagnetic shielding and antistatic coatings.

**Experimental Part**
**Materials**
Graphene oxide (GO, Cheaptubes USA). The acrylic resin polyethyleneglycol diacrylate (PEGDA, Ebecryl 11 Cytec Belgium, Mw ≈ 740 g/mol, density = 1.12 g/cm$^3$). The radical photoinitiator 1-[4-(2-idrossietossi)fenil]-2-idrossi-2-metil-1-propan-1-one (Irgacure 2959, Ciba).

**Coatings preparation**
GO was dispersed in water (concentration 10 mg/ml) by Ultraturrax mixing. The water dispersion was added to PEGDA at a content in the range from 0.5 to 2 % by weight. The formulations were added with 2 % by weight of the radical photoinitiator with respect to PEGDA, coated on glass slides and cured, under nitrogen, for one minute with a light intensity on the surface of the sample of 30 mW/cm$^2$. PEGDA-GO photocured nanocomposites were characterized by a final GO content varying in the range 0.005 - 0.2 % wt, respectively.

**Characterization**

Acrylic double bond conversion, as a function of irradiation time, was evaluated by means of real-time FT-IR spectroscopy, employing a Thermo-Nicolet 5700 instrument. For this purpose, the formulations were coated onto silicon wafers and simultaneously exposed to the UV beam, which induces polymerization, and to the IR beam, which makes an in-situ evaluation of the extent of reaction. Acrylic double bond conversion was followed by monitoring the decrease in the absorbance at 1450 $cm^{-1}$ due to C=C double bonds (the acrylic double bond conversion was normalized with respect to the carbonyl peak centred at 1700 $cm^{-1}$). A medium pressure mercury lamp (Hamamatsu) equipped with an optical guide was used to induce photopolymerization (light intensity on the surface of the sample of about 30 $mW/cm^2$).

The gel content was determined on the cured films by measuring the weight loss after 24 hours extraction in chloroform at room temperature, according to the standard test method ASTM D2765-84.

UV-Vis measurements of the GO-water dispersion formulation or the casted films were carried out with a Perkin-Elmer spectrometer Lambda 35; for all samples, a quartz slide was used as the reference.

Morphology of the polymer composites and graphene dispersion were investigated by optical microscopy, field emission scanning electron microscopy (FE-SEM) and transmission electron microscopy (TEM, JEOL). For FE-SEM analyses the surface fracture of the cured coatings were observed with the In-Lens detector. This detector is an ideal tool to investigate polymeric materials thanks to its high detection efficiency at very low acceleration voltages and the almost pure detection of SE electrons. The detector is placed above the objective lens and detects directly the beam path. The lower the energy of the primary electrons, the smaller the interaction volume and the penetration depth of the electrons will be. The smaller penetration depth of the electrons, the higher the share of SE electrons generated in the upper layers of the specimen, which contribute to the image contrast and resolution. This detector allows to collect images at very low acceleration voltages (1.5-5 kV) with the minimization and compensation of the effects due to the accumulation of local charges on the surface of non-conductive materials, that otherwise can significantly deteriorate the imaging quality.

Samples for TEM were photocured directly onto copper grids (150 mesh), after deposition of a small amount of graphene loaded PEGDA formulations.

DSC measurements were performed under nitrogen flux, in the range between -20 °C to 100 °C, with a DSCQ 1000 of TA Instruments equipped with a low temperature probe.

The electrical characterization (i. e. four probe measurements and I-V characteristic) was performed using a computer-controlled Keithley 4200 Source Measure Unit. For the four probe measurements, Al-electrodes (1 mm × 10 mm) having an average thickness of ~70 nm thermally evaporated on the PEGDA/GO coated quartz substrates maintaining a spacing of 2 mm were used.

**Results and Discussion**

Water dispersion of GO was prepared by Ultraturrax mixing, and added into PEGDA as acrylic UV-curable system in the range between 0.5 to 2 % by weight with respect to PEGDA; this corresponds to a final concentration of GO in PEGDA as low as 0.005 – 0.02 % by weight in the composite, respectively. The morphology of the GO-water dispersion was previously evaluated using optical microscopy, FE-SEM and AFM techniques. At eye view, the liquid dispersion appeared coloured but completely transparent and stable for several days. Microscopic investigation (optical microscopy and FE-SEM) showed that GO flakes with a thickness of 1-2 nm underwent exfoliation in water after Ultraturrax mixing, and that the dispersion contained extended flakes with a mean lateral dimension of 30-40 μm (see Figure 1 a-c). Addition of GO-water dispersions to the acrylic resin was pursued since the direct dispersion of GO into the resin induces immediate aggregation. In this regard it was previously reported [25] that the X-ray diffraction patterns for samples obtained from GO-water dispersion prepared under analogous experimental conditions show the absence of any diffraction peaks. The transparency of the pristine PEGDA photocurable formulations is maintained even after the addition of the GO-water dispersion.

The effect of the presence of GO-water dispersion on the photopolymerization process was investigated by means of real-time FT-IR. The conversion curves as a function of irradiation time for the pristine PEGDA resin and the PEGDA-GO (water dispersion) system were recorded and reported in Figure 2: in the curves, the plateau value gives the final acrylic double bond conversion, while the polymerization rate can be derived by the slope of the curve. The kinetic curves of the pristine PEGDA resin showed a high rate of polymerization and a double bond conversion of about 70% after one minute of irradiation. The addition of GO-water dispersion up to a content of 2 % by weight, did not significantly influence the rate of UV-curing and the final acrylic double bond conversion. It must be noted that, for reasons due to the experimental set-up, the real-time conversion curves were collected in air; in this conditions oxygen inhibits the radical processes, thus limiting the maximum conversion of the double bonds that can be reached. Definitely higher conversion values should be expected by performing analogous experiments under a nitrogen atmosphere, most likely reaching the complete conversion of acrylic double bonds. In Figure 3 the FT-IR spectra before (curve b, blue colour) and after (curve a, red colour) one minute of UV irradiation of pristine PEGDA are reported. The almost complete disappearance of the peak centred at 1450 cm$^{-1}$, indicating an almost complete conversion of acrylic double bonds, can be easily observed. In view of potential applications in the coatings technology, it is worth noting that the presence of GO does not significantly alter the FT-IR spectrum of the acrylic resin and its route of polymerization.

High gel content values (above 98%, see Table 1) were measured for the all the cured films, as an indication of the formation of a tight crosslinked network and the absence of extractable oligomers in the cured system. This is in accordance with the FT-IR data and reinforces the evidence that GO-water dispersion does not influence the radical proces of curing.

After UV-curing the coatings showed a light brown colour (see digital photograph, Fig. 4a) and a substantial optical transparency in the visible range, evaluated by UV-VIS spectrometry at 550 nm [26,27]. Transparency was maintained also for the UV cured PEGDA nanocomposites containing 0.02 %wt of GO (transmittance>80%, Fig. 4b).

The morphology of the cured PEGDA systems containing increasing amounts of GO was firstly evaluated by optical microscopy (Fig. 5 a-c). The optical images show that graphene sheets underwent exfoliation in PEGDA and that the PEGDA polymer matrix contains GO platelets. The morphology of the films with 0.005 to 0.02 %wt of GO into PEGDA resin presented flakes (Fig. 5a) and a more extended networking structure with increasing the GO content (Figs. 5b-c). From the optical transparency it can be deduced that the thickness of GO platelets dispersed into the resin is essentially the same of that measured for the neat GO dispersed in water. Once photopolymerized, the GO flakes undergo excellent conformal filling inside the PEGDA matrix. FE-SEM analysis as reported in Fig. 6, captured on the cross section of the fracture surface of the PEGDA-GO nanocomposites, showed that graphene sheets retain their original shape and structure when included in the crosslinked polymer matrix. In particular, no aggregation of GO flakes has been recorded within the composite structure. TEM investigations (Fig.7) revealed the presence of a finer carbon nanostructure perfectly soaked in the acrylic matrix, that creates a continuous carbon network inside the polymer film.

Differential scanning calorimetry (DSC) was performed on cured films in order to evaluate the glass transition temperature ($T_g$) of the GO nanocomposites with respect to pristine PEGDA, that showes its $T_g$ at - 45 °C. It was found that the addition of GO to PEGDA did not significantly alter the $T_g$ of the cured films, due to the small amount of the filler that is unable to influence the mobility of the polymer chains (see Tab. 1).

Electrical characterization was performed on the thin GO-containing PEGDA coatings, in order to evaluate the effect of GO sheets on the conductivity of the polymer matrix at loading ratios as low as those used in this study (namely 0.005-0.02 %wt of GO).

The photocured PEGDA-GO system containing 0.02 %wt of GO showed a sheet resistances of about 6300 Ω/sq, and lowering the GO content to 0.01 and 0.005 %wt takes to sheet resistance values of $5*10^6$ Ω/sq and to an insulating film, respectively (Table 1). The current-voltage characteristic of such samples shows a semimetallic behaviour (see Supporting Information). These results clearly show that by it is possible to achieve conductive coatings by adding 0.02 %wt of GO into acrylic UV-curable resin.

It is known for GO being an insulating material [28] thus to explain the increase of the conductivity of PEGDA resin by graphene oxide addition, two aspects have to be monitored: the contribution of water to the surface conductivity when GO-water dispersion has been added to the resin and UV induced change on the GO surface functional groups. From our measurements we demonstrated that the addition of water does not alter the conductivity of the neat polymer, since photocuring of PEGDA in the presence of water without GO added gave rise to insulating materials, while the exposition to UV light changes the GO conductivity (see Supporting Information) rendering it conductive due to the loss of oxygen-containing functional groups [29].

At this stage it should be stressed that the higher conductivity of the films obtained from such dispersion is not consistent with the electrical insulating behaviour of GO flakes; in fact the most convenient method for producing conductive graphene based composites involves the reduction of pre-fabricated graphene oxide sheets.[30,31] Moreover if we compare our results to those obtained by Eda et al. and Wu et al. which produced graphene films that had $R_s \approx 10^5$ Ω/sq [32] and $R_s \approx 3500$ Ω/sq [33] respectively, we could speculate that during the UV curing process the oxygen functionalities have been reduced in the graphene basal plane, possibly by the UV-generated radicals. Deeper investigation of this mechanism is currently under progress and will be reported in a future paper.

**CONCLUSION**

In this paper we report an easy route towards the development of transparent conductive graphene oxide (GO)-acrylic resin coatings achieved by photopolimerization. GO undergoes exfoliation in water after Ultraturrax mixing, and the homogeneous GO-water dispersion can be easily added into PEGDA resin without invalidation of PEGDA transparency. The addition of the GO-water dispersion up to a content of 2 wt% with respect to PEGDA does not significantly influence the rate of UV-curing and the final acrylic double bond conversion. PEGDA-GO nanocomposite coatings present a good dispersion of graphene oxide sheets inside the polymer matrix: GO sheets maintain the exfoliated state reached in the water dispersion even when mixed with PEGDA. After photopolymerization, the GO flakes undergo excellent conformal filling within the PEGDA crosslinked matrix. Electrical conductivity measurements demonstrate that a conductive PEGDA coating can be easily obtained by this method by addition of amounts of GO as low as 0.02 %wt.

These new functional materials can find applications as transparent conductive thin films. Following these indications we retain that the presented synthesis strategy could have implications toward building some novel architectures with graphene and polymers.

Table 1. Gel content, glass transition temperatures, and surface resistivity ($R_s$) values for UV-cured acrylic coatings.

| Sample | Gel Content %[a] | $T_g$ °C[b] | $R_s$ Ω/sq |
|---|---|---|---|
| PEGDA | 98 | -45 | -*) |
| PEGDA+0.5 wt% GO | 98 | -45 | -*) |
| PEGDA+1 wt% GO | 97 | -45 | $5*10^6$ |
| PEGDA+2 wt% GO | 98 | -45 | 6300 |

a) Measured after 24 hours extraction in chloroform, ASTM D2765-84.

b) Measured by DSC analyses.

c) Four probe measurements surface resistance.

*) Surface resistance values higher than $10^{11}$ Ω/sq.

Figure 1: (a) Vial containing 0.1mg/ml dispersion of GO in water; (b) FE-SEM image of the extended GO flake obtained from water dispersion and (c) optical micrographs (400 μm X 200 μm) of GO flakes obtained from water dispersion.

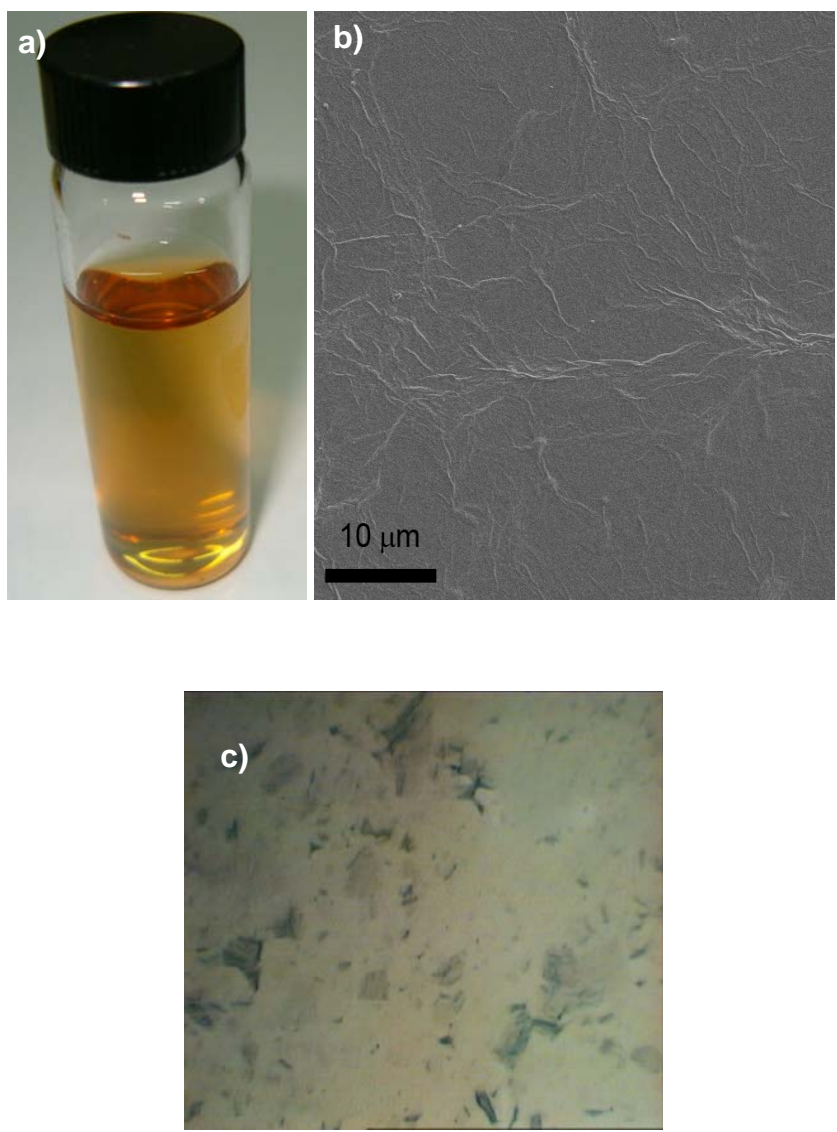

Figure 2: RT-FT-IR conversion curves as a function of irradiation time for pristine PEGDA resin and for formulations of PEGDA resin containing 0.5 wt% and 2.0 wt% of GO-water dispersion. Radical photoinitiator concentration 2 wt%. Light intensity 30 mW/cm$^2$. Film thickness 50 μm.

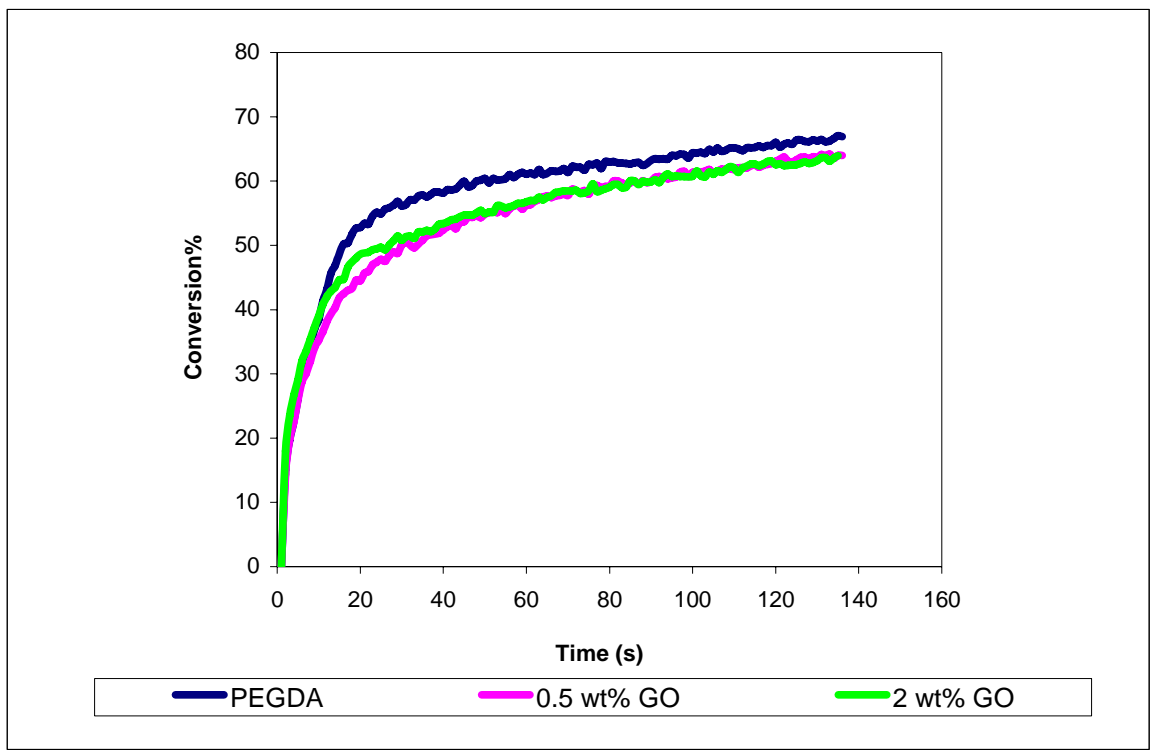

Figure 3: FT-IR before (red curve) and after (blue curve) one minute of irradiation of pristine PEGDA. Radical photoinitiator concentration 2 wt%. Light intensity 30 mW/cm$^2$. Film thickness 50 μm.

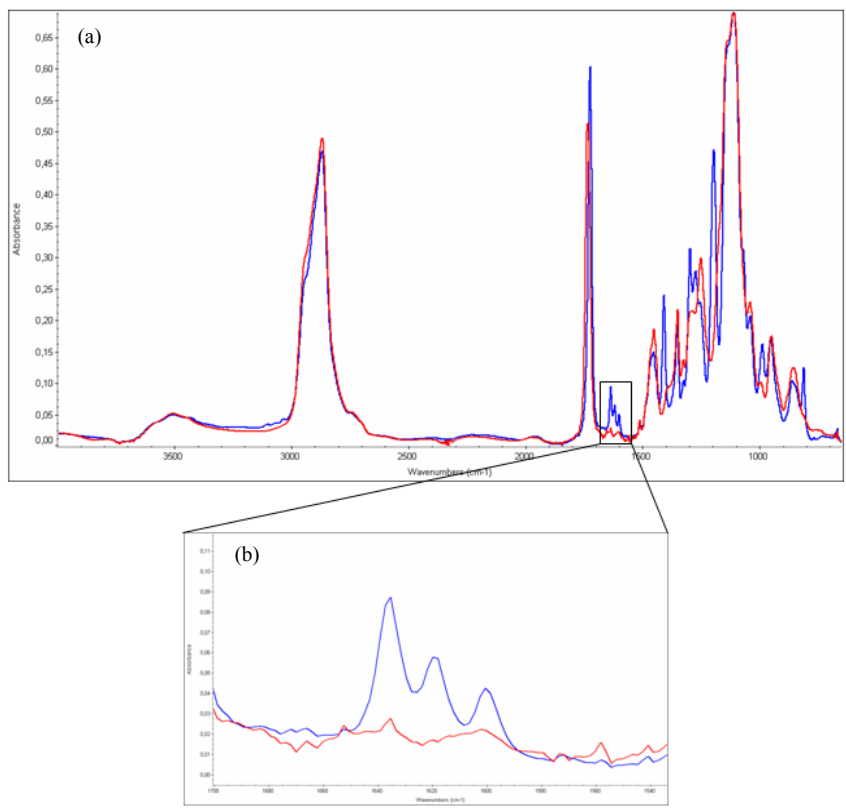

Figure 4: (a) Digital photograph of cured system obtained by adding 2 wt% of GO-water dispersion into PEGDA resin and (b) UV-Vis spectra of cured system obtained by adding 0,5 wt% (curve 1, black) and 2 wt% (curve 2, red) of GO-water dispersion into PEGDA resin.

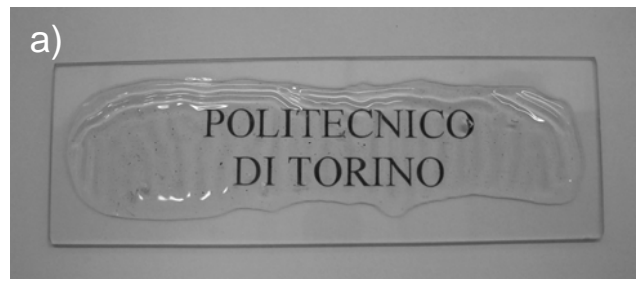

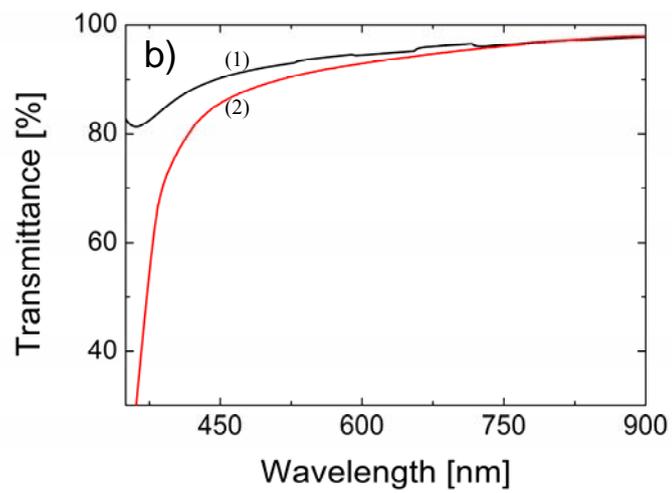

Figure 5: Optical micrographs (400 μm X 200 μm) of cured systems obtained by adding (a) 0,5 wt%, (b) 1 wt% and (c) 2 wt% of GO-water dispersion into PEGDA resin

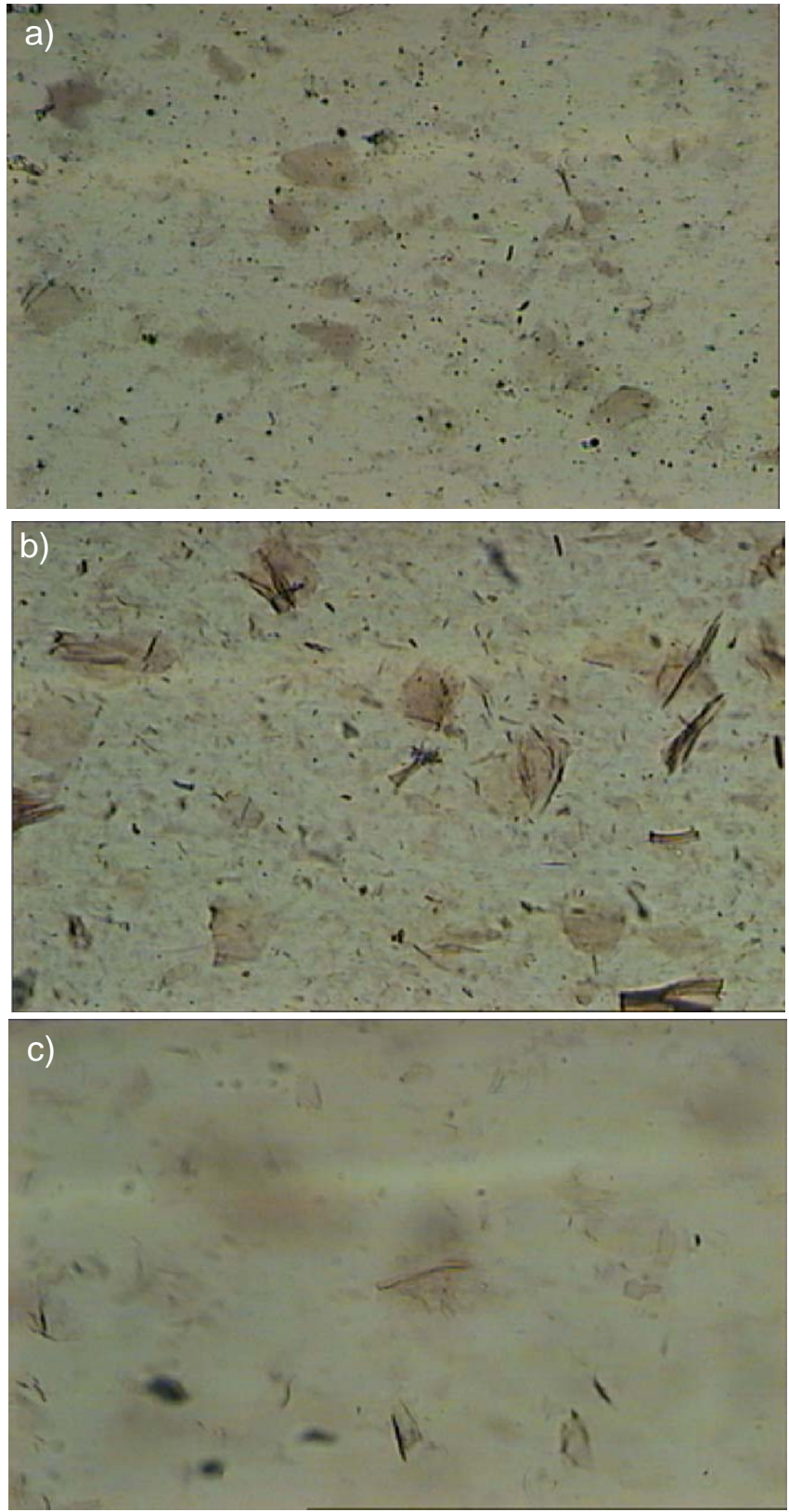

Figure 6: FE-SEM image of the fracture surface of the cured systems obtained by adding 2 wt% of GO-water dispersion into PEGDA resin

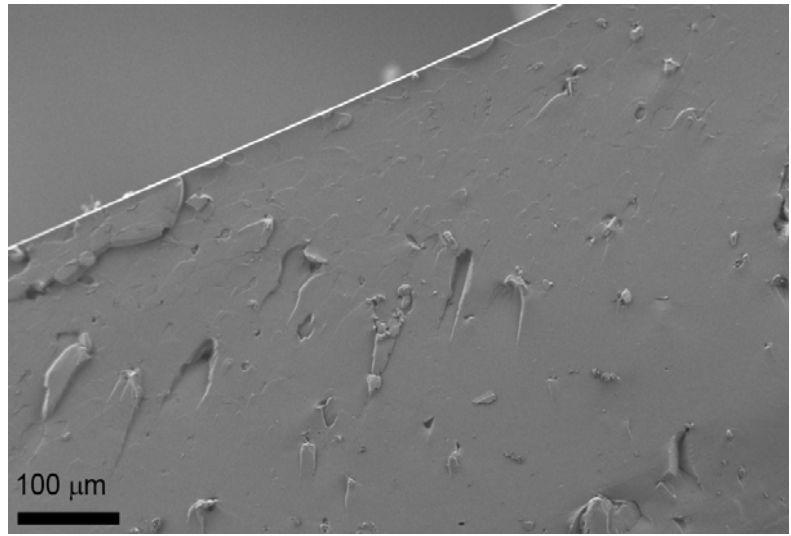

Fig.7: TEM image of the PEGDA composites with 1 wt% of GO dispersion.

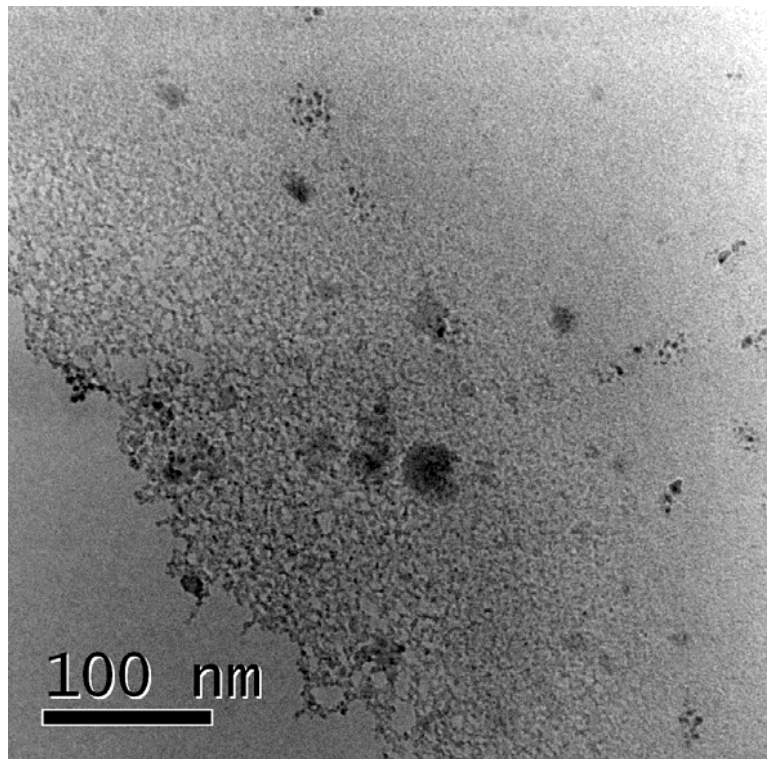